\documentclass[12pt]{article}
\usepackage{hyperref,amsfonts,amssymb, amsmath,theorem,mathrsfs}
\usepackage[english]{babel}
\usepackage[all]{xy}

\DeclareMathAlphabet{\mathpzc}{OT1}{pzc}{m}{it}

\def\bq{ \begin{equation} }
\def\eq{ \end{equation} }
\def\ben{ \begin{eqnarray} }
\def\en{ \end{eqnarray} }

\newtheorem{prop}{Proposition}

\theorembodyfont{\rm}

\begin{document}

\title{On the Steklov-Lyapunov case of the rigid body motion.}
\author{
A.V. Tsiganov\\
\\
\it\small  St.Petersburg State University, St.Petersburg, Russia}

 \date{}
\maketitle

\begin{abstract} {\small
We construct a Poisson map between manifolds with linear Poisson
brackets corresponding to the two samples of Lie algebra $e(3)$.
Using this map we establish equivalence of the Steklov-Lyapunov
system and the motion of a particle on the surface of the sphere
under the influence of the fourth order potential. To study
separation of variables for the Steklov case on the Lie algebra
$so(4)$ we use the twisted Poisson map between the bi-Hamiltonian
manifolds $e(3)$ and $so(4)$.}
\end{abstract}

\par\noindent
\textbf{Corresponding Author}:\\
A V Tsiganov, St.Petersburg State University, St.Petersburg, Russia\\
E-mail: tsiganov@mph.phys.spbu.ru \newpage
\section{Introduction}
 \setcounter{equation}{0}
A standard form of the Kirchhoff equations is the following:
\begin{equation}\label{Kirh-Eq}
 \dot{M}=M \times
\Omega+p \times U, \qquad \dot{p}=p \times \Omega,
\end{equation}
where $X\times Y$ stands for the vector product of
three-dimensional vectors \cite{kirh74}.

These equations describe the motion of a rigid body in the ideal
incompressible fluid if two vectors $M $ and $p$ are the impulsive
momentum and the impulsive force while the vectors $\Omega$ and
$U$ are the angular and linear velocities of the body. All these
vectors in ${\mathbb R}^3$ are expressed in the body frame
attached to the body \cite{kirh74}.

It is known  \cite{nov81} that the system (\ref{Kirh-Eq}) is
Hamiltonian with respect to the Lie--Poisson bracket of the Lie
algebra e(3) of the Lie group $E(3)$ of Euclidean motions of
$\mathbb R^3$ , i.e. with respect to the Poisson bracket
\bq\label{new-e3} \{M_{i}, M_{j}\}_1=\varepsilon_{ijk}M_{k}, \quad
\{M_{i}, p_{j}\}_1 =\varepsilon_{ijk}p_{k}, \quad \{p_{i},
p_{j}\}_1=0\,,
\eq
where $\varepsilon_{ijk}$ is the totally skew-symmetric tensor.
Here and below we identify dual space $e^*(3)$ with $e(3)$ using
the standard inner product \cite{nov81} .

The Hamiltonian equations of motion for an arbitrary Hamilton
function $H = H (p,M)$ in the bracket (\ref{new-e3}) read
\begin{equation}\label{Kirh-Eqh}
\dot{M}=M \times \nabla_M H+p \times \nabla_p H, \qquad \dot{p}=p
\times \nabla_M H.
\end{equation}
This Euler's equations on $e^*(3)$ coincides with (\ref{Kirh-Eq})
if
\[H(p,M) = \langle M,\mathbf A_1M\rangle
+\langle M, \mathbf A_2 p\rangle+\langle p,\mathbf A_3 p\rangle.\]
Here $\mathbf A_k$ are special numerical matrices and $\langle
.,.\rangle$ stands for the standard scalar product in $\mathbb
R^3$ \cite{nov81}.

The Poisson bracket (\ref{new-e3}) has two Casimir function
\bq\label{caz1}
\mathcal A=|p|^2,\qquad \mathcal B=\langle p, M \rangle \,.
\eq
 which are therefore integrals of motion for (\ref{Kirh-Eq}) in involution with Hamiltonian $H(p,M)$ and with any other function on the phase space.

The Steklov-Lyapunov case of the rigid body motion (the
Steklov-Lyapunov system, for brevity), is characterized by the
following diagonal matrices
\bq\label{AC-mat}
\mathbf A=\mbox{\rm diag}(a_1,a_2,a_3),\qquad \mathbf C=\mbox{\rm
diag}(a_2-a_3,a_3-a_1,a_1-a_2)\,.
\eq
In \cite{stek93} Steklov found a Hamilton function
\bq \label{H-St}
4H_S(p,M)=\langle M,\mathbf A M\rangle+2\langle M,\mathbf A^\vee
p\rangle+\langle \mathbf A p,\mathbf C^{\,2} p\rangle\,
\eq
for which equations (\ref{Kirh-Eqh}) possess a fourth additional
integral. Here wedge denotes adjoint matrix, i.e. cofactor matrix.
In our case it reads $\mathbf A^{\vee}=({\det \mathbf A})\,\mathbf
A^{-1}$.

Later Lyapunov \cite{lyap97} independently discovered an
integrable case of the Kirchhoff equations whose Hamiltonian
\bq\label{H-L}
4H_L(p,M)=\langle M,M \rangle-2\langle M,\mathbf A p \rangle+\langle
p,\mathbf C^{\,2} p\rangle
\eq
is a linear combination of the Steklov integrals \cite{stek93} and
the Casimir functions (\ref{caz1}).

The Lax matrices for the Steklov-Lyapunov system may be extracted
from the K\"otter work \cite{bob83, bbe94}. Namely, in
\cite{kot00} K\"otter introduce two vectors $\ell$ and $m$
\bq\label{vLax-K}
\ell(\lambda)=\mathbf W\left(\dfrac{1}{2}(M-\mathbf B p)
+\lambda\,p\right)\,,\qquad m(\lambda)=\mathbf W^\vee p\,,
\eq
where $\mathbf W$ and $\mathbf B$ are diagonal matrices with the
following entries
\bq\label{WB-mat}
\mathbf W_{ii}=\sqrt{\lambda-a_i},\qquad \mathbf
B_{ii}=\sum_{j,k=1}^{n=3} |\varepsilon_{ijk}|\,a_k\,.
\eq
Here $\lambda$ is auxiliary variable (spectral parameter) and
functions $\mathbf W_{ii}=\sqrt{\lambda-a_i}$    can be considered
as basic elliptic functions  (see \cite{bob83,bbe94}). The
equations of motion
\bq\label{st-eqm}
\dfrac{d}{dt}{\ell}(\lambda)\equiv\{H_L,\ell\}_1= m(\lambda)\times
\ell(\lambda)\,,
\eq
may be rewritten in the Lax form
\bq\label{Lax-Eq}
\dfrac{d}{dt}{\mathscr L}_e(\lambda)=\left[{\mathscr
M}_e(\lambda),{\mathscr L}_e(\lambda)\right]
\eq
using two Lax matrices
\bq\label{Lax-K}
{\mathscr L}_e(\lambda)=\sum_{i=1}^3 \ell_i(\lambda)\sigma_i\,,
\qquad {\mathscr M}_e(\lambda)=\sum_{i=1}^3
m_i(\lambda)\sigma_i\,,
\eq
where $\sigma_i$ are the  Pauli matrices.

For an actual integration of the corresponding Hamiltonian flow in
terms of elliptic functions see \cite{kot00}, and for a more
modern account \cite{bob83, bbe94}. In fact in the known
integration procedure we have to use the Lax matrices
(\ref{Lax-K}) with the spectral parameter $\lambda$ varying on an
elliptic curve.

However, it is known that the Steklov-Lyapunov flow is
linearizable on the Jacobian of hyperelliptic curve instead of
elliptic curve, similar to the Neumann system  \cite{bbe94}.
Therefore, we can suppose that the Steklov-Lyapunov system belong
to the family of the St\"ackel systems \cite{ts99} and there are
the Lax matrices with rational dependence on the spectral
parameter.

The desired rational Lax matrices were constructed by Bolsinov and
Fedorov by using a special triplets of vectors, which are
coordinates on some complicated phase space (see \cite{fed03} and
references within).

The aim of this note is to identify the Steklov-Lyapunov system
with an integrable motion of a particle on the surface of the
sphere, which is the St\"ackel system. This result is the direct
sequence of the K\"otter separation of variables \cite{kot00}.

\section{Separation of variables}
 \setcounter{equation}{0}

If $\sigma_k$ are $3\times 3$ Pauli matrices, then the Lax matrix
${\mathscr L}_e(\lambda)$ (\ref{Lax-K}) looks like
\[
{\mathscr L}_e(\lambda)=\left(
\begin{array}{ccc}
0& \ell_3& -\ell_2\\
-\ell_3& 0& \ell_1\\
\ell_2& -\ell_1& 0
\end{array}
\right)\in so(3)\,.
\]
The corresponding spectral curve  is defined by equation
$\det(\mu- \mathscr L_e(\lambda) )=0$, which is reduced to the
following equation
\bq\label{curve-Ell}
 \mathcal C:\qquad\mu^2+P_3(\lambda)=0\,,
\eq
where
\bq\label{P3-K}
P_3(\lambda)=|\ell|^2=\alpha^2\lambda^3+(\beta-\alpha^2\mbox{\rm
tr}\,\mathbf A)\lambda^2+{H_1}\,\lambda+{H_2}\,.
\eq
Coefficients  $H_{1,2}$ are linear combination of the Steklov-
Lyapunov integrals (\ref{H-St}-\ref{H-L}) and the Casimir
functions (\ref{caz1})
\bq\label{H12-SL}
H_1=H_L-\dfrac{\mathcal B}2\,\mbox{\rm tr}\,\mathbf A +\mathcal
A\,\mbox{\rm tr}\,\mathbf A^\vee\,,\qquad H_2=-H_S +\dfrac{\mathcal
B}2\,\mbox{\rm tr}\,\mathbf A^\vee-\mathcal A\,\det\,\mathbf A\,.
\eq

The separation of variables associated with the curve
(\ref{curve-Ell}) was constructed by K\"otter \cite{kot00} and may
be recovered in framework of the modern  Sklyanin method
\cite{skl95}.

\begin{prop} Separated variables $u_{1,2}$ associated with the
Lax matrix ${\mathscr L}_e(\lambda)$ are poles of the
corresponding Baker-Akhiezer function $\boldsymbol \Psi(\lambda)$
with the following dynamical nor\-ma\-li\-za\-ti\-on
\bq\label{norm}
\boldsymbol \alpha=\dfrac{a}{|(M-\mathbf B p)\times p\,|}\,\mathbf
W\,p,\qquad a\in \mathbf R\,.
\eq
\end{prop}
The proof consists of direct comparison of the known separated
variables \cite{kot00} with poles of the Baker-Akhiezer function
$\boldsymbol \Psi(\lambda)$, which is  an eigenvector of the Lax
matrix
\bq\label{BA-fun}
{\mathscr L}_e(\lambda)\boldsymbol \Psi(\lambda)=\mu\boldsymbol
\Psi(\lambda)\,.
\eq
Since an eigenvector is defined up to a scalar factor one has to
fix a normalization of $\boldsymbol \Psi(\lambda)$ imposing a
linear constraint
\bq\label{BA-norm}
\langle\boldsymbol \alpha, \boldsymbol \Psi(\lambda)\rangle=1.
\eq
In our case ${\mathscr L}_e(\lambda)\in so(3)$ and  excluding
$\mu$ from (\ref{BA-fun}-\ref{BA-norm}) we derive that poles
$u_{k}$ of $\boldsymbol \Psi(\lambda)$ are roots of the following
equation
\[
\bigl\langle\boldsymbol \alpha\times \ell,\boldsymbol \alpha\times
\ell\bigr\rangle=0\,,
\]
where $\ell$ is the K\"otter vector (\ref{vLax-K}).

 Inserting $\boldsymbol \alpha$
(\ref{norm}) in this equations and dividing it on polynomial
$\det(\lambda -\mathbf A)$ one can  define the separated variables
$u_{1,2}$ as roots of the  following function
\bq\label{sep-K}
e(\lambda)= \dfrac{(\lambda-u_1)(\lambda-u_2)}{\det(\lambda
-\mathbf A)}=\dfrac{\langle\boldsymbol \alpha\times
\ell,\boldsymbol \alpha\times \ell\rangle}{\det(\lambda -\mathbf
A)}
\equiv\sum_{i=1}^{n=3}\dfrac{x_i^2}{\lambda-a_i}=0\,,
\eq
where vector $x$ is given by
\[
x=a\dfrac{(M-\mathbf B p)\times  p}{|(M-\mathbf B p)\times
p\,|}\,,\qquad a\in \mathbb R\,.
\]
The equation (\ref{sep-K}) coincides  with  definition of the
separated variables from \cite{kot00}, where  K\"otter also proved
that initial equations of motion (\ref{Kirh-Eq}) can be written in
the form
\bq\label{evsep-K}
\dot{u}_1=\dfrac{2\sqrt{P_3(u_1)\det(\mathbf A-u_1
)\,}}{u_1-u_2},\qquad \dot{u}_2=-\dfrac{2\sqrt{P_3(u_2)\det(\mathbf
A-u_2)\,}}{u_1-u_2}
\eq
and then he integrated these equations using Abel-Jacobi inversion
theorem.

\begin{prop}
The K\"otter variables $u_{1,2}$ (\ref{sep-K}) and   momenta
$v_{1,2}$ defined by
\bq\label{v-K}
v_j=\left.
-\dfrac{1}{2a^2}\,\{H_1,e(\lambda)\}\right|_{\lambda=u_j}\qquad
j=1,2,
\eq
are canonical variables
\[ \{u_1,u_2\}_1=\{v_1,v_2\}_1=0,\qquad
\{v_j,u_k\}_1=\delta_{jk}\,,
\]
which satisfy to the following separated equations
\bq\label{sepeq-K}
v_j^2+\dfrac{P_3(u_j)}{\det(u_j-\mathbf A)}=0\,,\qquad j=1,2.
\eq
\end{prop}
The proof is straightforward.

As sequence of (\ref{sepeq-K}), in canonical variables $u_{1,2}$ and
$v_{1,2}$ integrals of motion $H_{1,2}$ are the St\"ackel integrals
\bq\label{int-St}
H_{j}=\sum_{k=1}^2 \mathcal
S^{-1}_{jk}\Bigl(\varphi(u_k)\,v_k^2-U(u_k)\Bigr)\,,\qquad
j=1,2,
\eq
where
\[
\varphi(\lambda)=\det(\mathbf A-\lambda)\,,\qquad
U(\lambda)=\alpha^2\lambda^3+(\beta-\alpha^2\mbox{\rm tr}\,\mathbf
A)\lambda^2\,,
\]
and $\mathcal S$ is the St\"ackel matrix
\bq\label{St-mat}
\mathcal S=\left(%
\begin{array}{cc}
  u_1 & u_2 \\
  1 & 1 \\
\end{array}%
\right)\,.
\eq
This matrix is $2\times 2$  block of the transpose Brill-Noether
matrix $\mathcal U_{\,\mathcal C}$
\bq\label{BN-mat}
\mathcal U_{\,\mathcal C}=\left(%
\begin{array}{cccc}
  u_1^3 & u_1^2 & u_1 & 1 \\
  u_2^3 & u_2^2 & u_2 & 1
\end{array}%
\right)\,,
\eq
which determines the Abel-Jacobi map on Jacobian of hyperelliptic
curve $\mathcal C$ (\ref{curve-Ell}) completely.

In our case the St\"ackel matrix $\mathcal S$ (\ref{St-mat}) is
the lowest block of the transpose Brill-Noether matrix $\mathcal
U_{\,\mathcal C}$ (\ref{BN-mat}))  and, therefore, there are
canonical coordinates in which equations of motion (\ref{evsep-K})
are the Newton equations \cite{ts99,ts99b}.

In the next section we discuss this canonical change of variables
$(p,M)\to (x,J)$ which transforms  the Kirchhoff equations
(\ref{Kirh-Eq}) to the Newton equations in detail.

\section{The Poisson map}
 \setcounter{equation}{0}
The Poisson manifold is a smooth manifold $\mathcal M$ endowed
with the Poisson brackets $\{.\,,.\}_{\mathcal M}$. If $\mathcal
M$ and $\mathcal N$ are Poisson manifolds, a smooth map
$f:\mathcal M\to \mathcal N$ is called  a Poisson map provided
that it preserves Poisson brackets, i.e.
\[f^*\Bigl\{\varphi,\psi\Bigr\}_{\mathcal
N}=\Bigl\{f^*\varphi,f^*\psi\Bigr\}_{\mathcal M}
\]
for all  $\varphi,\psi\in C^{\infty}(\mathcal N)$. Here
$f^*\varphi=\varphi\circ f$ is a lifting of the function
$\varphi\in C^{\infty}(\mathcal N)$ on $\mathcal M$.

Below we deal with linear Poisson brackets corresponding to the
two samples of Lie algebra $e(3)$ and to one sample of $so(4)$
algebra, i.e. with homomorphisms of these Lie algebras. For
brevity we will use the same notations both for the Poisson
manifolds and the Lie algebras.

If ${\mathcal M}$ coincides with ${\mathcal N}$, the Poisson maps
are called {\it canonical transformations}. The problem of
complete efficient description of all nonlinear canonical
transformations is unsolvable. The reason is that for any function
$f(\mathbf x_1,\ldots,x_n)$ the flow defined by ODEs $
\dot{x}_i=\{f\,,x_1,\ldots,x_n\}$ yields a one-parameter group of
canonical transformations. However one can investigate some
interesting subgroups of nonlinear canonical transformations.

Let $x$ and $J$ are coordinates on the Lie algebra $e(3)$ with the
standard Lie-Poisson brackets
\begin{equation}\label{e3}
\,\qquad \bigl\{J_i\,,J_j\,\bigr\}'_1=\varepsilon_{ijk}J_k\,,
\qquad \bigl\{J_i\,,x_j\,\bigr\}'_1=\varepsilon_{ijk}x_k \,,
\qquad \bigl\{x_i\,,x_j\,\bigr\}'_1=0\,.
\end{equation}
The brackets (\ref{e3}) respect two Casimir elements
\begin{equation}\label{caz0}
A=|x|^2\equiv\sum_{k=1}^{n=3} x_k^2, \qquad
   B=\langle x, J\rangle\equiv\sum_{k=1}^{n=3}  x_kJ_k \,.
\end{equation}
Fixing values of the Casimir elements one gets a generic
symplectic leaf of $e(3)$
\bq\label{symp-e3}
{\mathcal O}_{ab}= \{x,J\,:~A=a^2,~~ B={b}\}\,,
\eq
which is topologically equivalent to cotangent bundle $T^*S^{2}$
of the sphere
\[S^{2}=\{x\in \mathbb R^3, |x|=a\}.\]
Symplectic structure of $\mathcal O_{ab}$ is different from the
standard symplectic structure on $T^*S^2$ by the magnetic term
proportional to $b$ \cite{nov81}.

If $b=0$ there is standard Poisson map $T^*S^{2}\to e(3)$
\bq\label{xp-J}
\rho:\quad (\pi,x)\to J=\pi\times  x, \eq where $\pi\in \mathbb
R^3$ is conjugated to $x$ momenta
\bq\label{xp-S}
\{\pi_i,x_j\}=\delta_{ij},\qquad\mbox{\rm and}\qquad \langle x,
\pi\rangle=0.\eq

Let us consider  classical counterpart of the Fourier
transformation  $f:T^*S^2\to T^*S^2 $ defined by
\bq\label{fs-map}
f:\quad (x,\pi)\to(-\pi,x)\,.
\eq
This symplectic mapping may be lifted to the Poisson mapping
$\widetilde{f}:e(3)\to e(3)$
\bq \label{fp-map}
\widetilde{f}:\quad (p,M)\to
\left(x=\sqrt{a^2-\dfrac{b^2}{|M|^2}\,}\,\dfrac{M\times
p}{|\,M\times p\,|}+b\dfrac{M}{|M|^2}\,,\qquad J=M\,\right).
\eq
Its inverse mapping looks like
\bq\label{inv-fp}
\widetilde{f}^{-1}:\quad (x,J)\to
\left(\,p=-\sqrt{\alpha^2-\dfrac{\beta^2}{|J|^2}\,}\,\dfrac{J\times
x}{|\,J\times x\,|}+\beta\dfrac{J}{|J|^2}\,,\qquad M=J\,\right).
\eq
The maps $\widetilde{f}$  and  $\widetilde{f}^{-1}$ couple
coordinates $(x,J)$ on algebra $e(3)$ (\ref{e3}) with the
following values of the Casimir functions (\ref{caz0})
\[A=a^2,\qquad B=b\]
and coordinates $(p,M)$ on another sample of $e(3)$ (\ref{new-e3})
with the following values of the Casimir elements (\ref{caz1})
\[\mathcal A=\alpha^2\,,\qquad \mathcal B=\beta.\]

The symplectic  map $f$ (\ref{fs-map}) on $T^*S^2$ is easy
generalized:
\[f_g:\qquad (x,\pi)\to(-\pi+g(x),x)\]
if $g(x)$ is a function on $x$ such that
\[\langle
x,g(x)\rangle=0.\] For instance we can put $g(x)=x \times \mathbf
Bx$, where $\mathbf B$ is an arbitrary numerical matrix.

For brevity the lifting of this symplectic map $f_g$ to the
Poisson map we present at the special case  $\mathbf B=\mbox{\rm
diag}(b_1,b_2,b_3)$ and $b=0$ only. It will be enough in order to
identify the Steklov-Lyapunov system with another integrable
systems on $e(3)$ \cite{ts04}.

\begin{prop}
Let $\mathbf B$ and $\mathbf C$ are  numerical diagonal  matrices
with the following entries
\bq\label{BC-mat}
\mathbf B_{ii}=b_i\,,\quad \mathbf C_{ii}=\sum_{j,k=1}^{n=3}
\varepsilon_{ijk}\, b_j\,,\qquad b_i\in \mathbb R.
\eq
The mapping $\widetilde{f}_g: (p,M)\to (x,J)$ defined by
\bq\label{fg}
x=a\dfrac{(M-\mathbf B p)\times  p}{|(M-\mathbf B p)\times
p\,|},\qquad
  J=M+a^{-2}\mathbf C \Bigl[\, x,x\times  p\,\Bigr]_+\,,
\eq
is a Poisson map $\widetilde{f}_g:e(3)\to  e(3)$ such that
\[
A=|x|^2=a^2,\qquad B=\langle x,J \rangle=0.
\]
Here $[y,z]_+$ is an "anticommutator" of  two vectors $y$ and $z$
defined by
\[{[y,z]_+}_i=\sum_{j,k=1}^{n=3}|\varepsilon_{ijk}\,| \,y_j\,z_k\,.
\]
\end{prop}
The proof is straightforward verification of the Poisson brackets.

Using  $\mathbf B$ and $\mathbf C$ (\ref{BC-mat}) we determine
matrix $ \mathbf A=\mbox{\rm diag}(a_1,a_2,a_3)$  and its adjoint
$\mathbf A^\vee$
\bq\label{A-mat}
\mathbf A=\dfrac12\mbox{\rm tr}\,(\mathbf B)\, -\mathbf
B\qquad\mbox{\rm and}\qquad \mathbf A^\vee=\dfrac14(\mathbf
C^2-\mathbf B^2)\,.
\eq
In these notations let us consider motion of a particle on the
surface of the unit sphere
\bq\label{u-S}
S^2=\{x\in \mathbb R^3,\,|x|=1\}
\eq
under an influence of the fourth order potential. The Hamilton
function is equal to
\bq\label{H1-S}
\widetilde{H}_1(x,J)=\dfrac14\langle J,J\,\rangle -\langle x, \mathbf
B x\rangle\Bigl(\gamma
 +\delta\langle x, \mathbf B x\rangle\Bigr)
 +\delta\langle x,\mathbf A^\vee x\rangle\,,
\eq
where $\gamma,\delta$ are arbitrary parameters. If $\delta=0$ the
function $\widetilde{H}_1(x,J)$ (\ref{H1-S}) is Hamiltonian for
the Neumann system. Recall that we can put $|x|=1$ without loss of
generality using canonical transformations $x\to a^{-1}x$.

\begin{prop}\label{prop-PM} Let $\alpha^2$ and $\beta$ are the values of the Casimir
elements $\mathcal A$ and $\mathcal B$ (\ref{caz1}) on $e(3)$
algebra (\ref{new-e3}). If
\bq\label{Usl-St}
\gamma=\beta-\alpha^2\mbox{\rm tr}\,(\mathbf A)\,,\qquad
\delta=\alpha^2
\eq
the Poisson map $\widetilde{f}_g$ (\ref{fg}) identifies  the
Hamilton function $\widetilde{H}_1$ (\ref{H1-S}) on $T^*S^2$ with
the following Hamiltonian on the algebra  $e(3)$ (\ref{new-e3})
\[H_1=\widetilde{f}_g\Bigl(\widetilde{H}_1\Bigr)=H_L-
\dfrac12\mathcal\beta\,\mbox{\rm tr}\,\mathbf A +\alpha^2\,\mbox{\rm
tr}\,\mathbf A^\vee \,.
\]
This Hamiltonian coincides with the Lyapunov integral $H_L$
(\ref{H-L}) up to the last two terms depending on the Casimir
elements.
\end{prop}
The proof is straightforward.

As a result we arrive at the following conclusion: the
Steklov-Lyapunov system is equivalent to the potential motion of a
point $x\in \mathbb R^3$ constrained to the sphere, which belongs
to the St\"ackel family of integrable systems \cite{ts04}.

For the uniform St\"ackel systems we know the Lax matrices, the
classical $r$-matrices, the bi-Hamiltonian description, the
B\"acklund transformations, the separated variables, the
theta-function solutions and many other facts. The Proposition
\ref{prop-PM} allows  to transfer all these know results
concerning to the St\"ackel systems onto the Steklov-Lyapunov
system directly .

\section{The rational Lax matrices}
It is known \cite{w85}, that the Hamiltonian
$\widetilde{H}_1(x,J)$ (\ref{H1-S}) is separable in elliptic
coordinates $\widetilde{u}_{1,2}$ on the unit sphere $S^2$
(\ref{u-S}), which are roots of the standard form
\bq\label{ell-S}
e(\lambda)=\dfrac{(\lambda-\widetilde{u}_1)(\lambda-\widetilde{u}_2)}{\det(\lambda
-\mathbf A)} =\langle x, (\lambda  -\mathbf A)^{-1}x \rangle\,.
\eq
According to \cite{mum84,ts99,ts99b} the generic  $2\times 2$ Lax
matrices for the uniform St\"ackel system are constructed using
Hamiltonian $H_1$ and the generating function $e(\lambda)$ of the
separated variables only
\bq\label{lax-G}
{\mathscr L}_r(\lambda)=\left(\begin{array}{lr}
-\dfrac{1}{2}\,e_t(\lambda)&e(\lambda)\\
\\
-\dfrac{1}{2}\,e_{tt}(\lambda)+ w(\lambda)\,e(\lambda)\quad&
\dfrac{1}{2}\,e_t(\lambda)\end{array}\right)\,,\qquad {\mathscr
A}_r(\lambda)=\left(\begin{array}{cc}
0&1\\
\\
w(\lambda)&0\end{array}\right)\,.
\eq
Here $e_t=\{H_1,e\}$ and  function  $w(\lambda)$ is given by $
w(\lambda)=\Bigl[\phi(\lambda)e(\lambda)^{-1}\Bigr]_{MN}$, where
$\phi(\lambda)$ is a parametric function  and $[\xi]_{MN}$ is the
linear combinations of the following Laurent  projections
\cite{ts99,ts99b}
\bq
{[ \xi ]_{N}}=\left[\sum_{k=-\infty}^{+\infty} z_k\lambda^k\,
\right]_{N}\equiv \sum_{k=-M}^{N} \xi_k\lambda^k\,.
\eq

In our case $N$=2, $M=0$ and
$\phi(\lambda)=-\delta\lambda-\gamma$. It's easy to prove that the
required Lax matrices (\ref{lax-G}) are given by
\bq\label{lax}
{\mathscr L}_r(\lambda)
=\sum_{i=1}^{n=3}\left(%
\begin{array}{cc}
  -\dfrac12\,\dfrac{x_i\pi_i}{\lambda-a_i} & \dfrac{x_i^2}{\lambda-a_i} \\
  -\dfrac14\,\dfrac{\pi_i^2}{\lambda-a_i} & \dfrac12\,\dfrac{x_i\pi_i}{\lambda-a_i} \\
\end{array}%
\right)-\left(\begin{array}{cc}
          0 & 0 \\ \\
          \delta\lambda+\delta\langle x,\mathbf B x \rangle+\gamma & 0 \\
        \end{array}\right)\,,
\eq
where  $x_i,\pi_i$  are  canonical coordinates (\ref{xp-S}) and
\[
{\mathscr A}_r(\lambda)=\left(\begin{array}{cc}
0&1\\
\\
-\bigl(\lambda-\langle x,\mathbf
Ax\rangle\bigr)\bigl(\delta\lambda+\delta\langle x,\mathbf B x
\rangle+\gamma\bigr)-\delta\bigl(\langle x,\mathbf A
x\rangle^2-\langle x,\mathbf A^2
x\rangle\bigr)&0\end{array}\right)\,.
\]
The Poisson bracket relations between the entries of the matrix
${\mathscr L}_r(\lambda)$ are closed into the standard linear
$r$-matrix algebra. The corresponding  $r$-matrix is rational
dynamical matrix \cite{ts99,ts99b}. Here we present only one relation
\bq\label{rmat-rel}
\left\{-\dfrac12\,e_t(\lambda),e(\mu)\right\}_1'
=\dfrac{e(\lambda)-e(\mu)}{\lambda-\mu}\,,
\eq which allows us to
introduce canonical momenta
\[\widetilde{v}_k=-\left.\dfrac12\,e_t(\lambda)\right|_{\lambda=\widetilde{u}_k}
=-\left.\dfrac12\,\sum_{i=1}^{n=3}\dfrac{x_i\pi_i}
{\lambda-a_i}\right|_{\lambda=\widetilde{u}_k}\,,
\]
such that $\{\widetilde{v}_k,\widetilde{u}_j\}=\delta_{kj}$.

Inserting  $\lambda=\widetilde{u}_k$ into the determinant of the Lax
matrix ${\mathscr L}_r(\lambda)$ (\ref{lax})
\[
\det{\mathscr
L}_r(\lambda)=\left(-\dfrac14e_t^2+\dfrac12e\,e_{tt}-w\,e^2\right)
=\dfrac{\widetilde{P}_3(\lambda)}{\det(\lambda -\mathbf A)}\] one
gets two equations
\[-\dfrac14e_t^2(\widetilde{u}_k)=
\dfrac{\widetilde{P}_3(\widetilde{u}_k)}{\det(\widetilde{u}_k
-\mathbf A)},\qquad k=1,2,
\]
where $\widetilde{P}_3$ is generating function of integrals of
motion
\bq\label{3P3}
\widetilde{P}_3(\lambda)=\delta\lambda^3+\gamma\lambda^2
+\lambda\widetilde{H}_1+\widetilde{H}_2.
\eq
In Hamiltonian variables $\widetilde{v}_{k},\,\widetilde{u}_{k}$
these equations are the separated equations
\bq\label{sep-Eq}
-\widetilde{v}_k^2=
\dfrac{\widetilde{P}_3(\widetilde{u}_k)}{\det(\widetilde{u}_k
-\mathbf A)}\,,\qquad k=1,2,
\eq
whereas in Lagrangian variables
$\dot{\widetilde{u}}_{k},\,\widetilde{u}_{k}$ they are equations
\[
-\dfrac14\left(\dfrac{(-1)^{k+1}
(\widetilde{u}_1-\widetilde{u}_2)\,\dot{\widetilde
{u}_k}}{\det(\widetilde{u}_k -\mathbf
A)}\right)^2=\dfrac{\widetilde{P}_3(\widetilde{u}_k)}{\det(\widetilde{u}_k
-\mathbf A)}\,,\qquad k=1,2,
\]
which may be integrated using Abel-Jacobi inversion theorem.

The second integral of motion  $\widetilde{H}_2$ in (\ref{3P3}) is
equal to
\bq\label{H2-S}
\widetilde{H}_2(x,J)=-\dfrac14\langle J,\mathbf A J\rangle +\langle
x,\mathbf A^\vee x \rangle\,\Bigl(\gamma +\delta\langle x,\mathbf B x
\rangle\Bigr)\,.
\eq
As functions on elliptic coordinates $\widetilde{u}_{1,2}$ and
momenta $\widetilde{v}_{1,2}$ integrals of motion
$\widetilde{H}_{1,2}$  (\ref{3P3}) on $T^*S^2$ coincide with the
St\"ackel integrals (\ref{int-St}).

So, if conditions (\ref{Usl-St}) hold then the Poisson map
$\widetilde{f}_g$ (\ref{fg}) identifies the second integral of
motion $\widetilde{H}_2$ (\ref{H2-S}) on $T^*S^2$ with the
following integral on the algebra $e(3)$ (\ref{new-e3})
\[
H_2(p,M)=\widetilde{f}_g\Bigl(\widetilde{H}_2\Bigr)=-H_S
+\dfrac12\beta\,\mbox{\rm tr}\,\mathbf A^\vee-\alpha^2
\,\det\,\mathbf A\,.
\]
This  is the  Steklov integral $H_S$ (\ref{H-St}) up to the
constant terms depending on the Casimir elements.

The same Poisson map $\widetilde{f}_g$ (\ref{fg}) identifies
standard elliptic coordinates $\widetilde{u}_{1,2}$ (\ref{ell-S})
on the sphere $S^2$ and the corresponding separated equations
(\ref{sep-Eq}) with the separated variables $u_{1,2}$
(\ref{sep-K}) and the separated equations (\ref{sepeq-K}) for the
Steklov-Lyapunov system proposed by K\"otter \cite{kot00}. It
allows us to use the standard finite-band integration technique
for the potential motion on the sphere \cite{mum84} in order to
verify the K\"otter solution of the Steklov-Lyapunov system in
terms of the theta-functions.

Applying the Poisson map  $\widetilde{f}_g$ (\ref{fg}) to the Lax
matrix ${\mathscr L}_r(\lambda)$ (\ref{lax}) one gets the rational
Lax matrix for the Steklov-Lyapunov system. In \cite{fed03} the
similar Lax matrix was constructed by using the triplet of the
vectors $x,y,v$ defined by
\[
(M-\textbf B\,p) =x\times y,\qquad p=x\times v\,.
\]

\section{The Steklov system on $so(4)$ and twisted Poisson map}
The bi-Hamiltonian manifold is a smooth manifold $\mathcal M$
endowed with a pair of compatible Poisson brackets
$\{.\,,.\}_{\mathcal M}$ and $\{.\,,.\}'_{\mathcal M}$. In
contrast with the Poisson manifolds we have two opportunities for
the action of the mapping  $f:\mathcal M\to \mathcal N$ preserving
both Poisson brackets
\[1.\quad f:
\begin{array}{c}
\xymatrix{ \{.\,,.\}_{\mathcal M}  \ar[r]  &
\ar[l]\{.\,,.\}_{\mathcal N}
\\
\{.\,,.\}'_{\mathcal M} \ar[r] & \ar[l]\{.\,,.\}'_{\mathcal N}
}\end{array}, \qquad 2.\quad f:\begin{array}{c}

\xymatrix{ \{.\,,.\}_{\mathcal M}  \ar[dr]  & \ar[dl]
\{.\,,.\}_{\mathcal N}
\\
\{.\,,.\}'_{\mathcal M} \ar[ur] & \ar[ul]\{.\,,.\}'_{\mathcal N}}
\end{array}.
\]
In order to distinguish these cases we will say about the Poisson
map and the twisted Poisson map at the first and the second cases
respectively.

Let us consider the Euler equations on the Lie algebra $so(4)$
\bq\label{PZ-eq}
\dot{s}=s \times \nabla_s  H(s,t), \qquad \dot{t}=t \times
\nabla_t
 H(s,t).
\eq
Here vectors $s$ and $t$ are coordinates on $so(4)=so(3)\oplus
so(3)$ with the standard Lie-Poisson brackets
\begin{equation} \label{2o3}
\bigl\{ s_i\,,s_j\,\bigr\}^\ast_1= \varepsilon_{ijk}\,s_k\,,
\qquad \bigl\{ s_i\,,t_j\,\bigr\}^\ast_1= 0\,,\qquad \bigl\{
t_i\,,t_j\,\bigr\}^\ast_1=\varepsilon_{ijk} t_k\,,
\end{equation}
The  brackets (\ref{2o3})  respect two Casimir  functions
\bq\label{caz4}
\mathcal A^*=|s|^2,\qquad \mathcal B^*=|t|^2 \,,
\eq
 which are therefore integrals of motion for (\ref{PZ-eq}) in
involution with  any function on the phase space.

Equations (\ref{PZ-eq}) describe the motion of a rigid body with
elliptic cavities filled with ideal fluid  if the Hamilton
function is  quadratic form
\[
H(s,t)=\langle s,\mathbf A_1s\rangle +\langle s, \mathbf A_2
t\rangle+\langle t,\mathbf A_3 t\rangle.\
\]
with the special numerical matrices  $\mathbf A_k$ \cite{poi01}.

In \cite{stek09} Steklov found the Hamilton function for which
equations (\ref{PZ-eq}) possess a fourth additional integral
\[H=c_1\widehat H_1+c_2\widehat H_2,\qquad
\left\{\widehat{H}_1,\widehat{H}_2\right\}^\ast_1=0\,,
\]
where $c_{1,2}$ are numerical parameters and
\bq\label{Int-St2}
 \widehat{H}_1=2\langle  \sqrt{\mathbf A^\vee}\, s,t\rangle-\langle
s,\mathbf A s\rangle\,,\qquad \widehat{H}_2=\langle t,\mathbf
A^\vee t\rangle-2\langle \sqrt{\mathbf A^\vee}\,s,\mathbf A t
\rangle\,.
\eq
In \cite{bob83} Bobenko established isomorphism between the
Steklov system on $so(4)$ and  the Steklov-Lyapunov system on
$e(3)$.
\begin{prop}\label{pr-bob} \cite{bob83}
If the  phase space $\widehat{M}\simeq so(4)$ is identified with
the space $M \simeq e(3)$ by the following linear map
\bq\label{st-pM}
\widehat{f}:\qquad (p,M)\to\left( s=2p,\quad t=
\,\dfrac{1}{\sqrt{\mathbf A^\vee}}\,\bigl(M-\mathbf B
p\bigl)\right)\,,
\eq
where $\mathbf A,\mathbf B$ are given by
(\ref{AC-mat},\ref{WB-mat}), then the equations of motion
(\ref{PZ-eq}) for the Steklov system on $so(4)$ coincide with the
Kirchhoff equations (\ref{Kirh-Eqh}) for the Steklov-Lyapunov
system on $e(3)$, that is,
\[
\bigl\{H_1,.\bigr\}_1=\bigl\{\widehat{H}_1,.\bigl\}^\ast_1\,.
\]
\end{prop}
Here we changed $p\to 2p$ and $M\to 2M$  in comparison with
original map \cite{bob83} to make formulas for the Poisson pencil
slightly more symmetric.

We have to underline that
\[\widehat{H}_1(s,t)\neq
\widehat{f}\Bigl(\,H_1(p,M)\Bigr)\,,
\]
as for the usual canonical or Poisson transformations. In fact the
map $\widehat{f}$ (\ref{st-pM}) gives rise the second Poisson
brackets on
 $so(4)$
\bq\label{sB-o4}
\{ s_i\,,s_j\,\bigr\}^\ast_2=0\,, \qquad \bigl\{
s_i\,,t_j\,\bigr\}^\ast_2=\varepsilon_{ijk}\,\dfrac{s_k}{\sqrt{\mathbf
A^\vee_{kk}}}\,,\qquad \bigl\{
t_i\,,t_j\,\bigr\}^\ast_2=\varepsilon_{ijk}
\left(\dfrac{t_k}{\mathbf A_{kk}}- \dfrac{s_k}{\sqrt{\mathbf
A^\vee_{kk}}}\right)\,,
\eq
whereas  inverse map $\widehat{f}^{-1}$ generates the second
Poisson brackets on $e(3)$
\bq \label{sB-e3}
\{M_{i}, M_{j}\}_2=\varepsilon_{ijk}\left(\mathbf
A_{kk}M_k-\dfrac{\mathbf C^\vee_{kk}}2p_k\right), \quad \{M_{i},
p_{j}\}_2 =\varepsilon_{ijk}\dfrac{\mathbf B_{kk}}2p_{k}, \quad
\{p_{i}, p_{j}\}_2=\varepsilon_{ijk} \dfrac{p_k}2\,.
\eq
Here $\mathbf C^\vee$ is the cofactor matrix to matrix $\mathbf C$
(\ref{AC-mat}).
\begin{prop}
The brackets $\{.,.\}_1$ (\ref{new-e3}) and $\{.,.\}_2$
(\ref{sB-e3}) on the manifold $e(3)$ and brackets $\{.,.\}^*_1$
(\ref{2o3}) and $\{.,.\}^*_2$ (\ref{sB-o4}) on the manifold
$so(4)$ are compatible.
\end{prop}
The proof consists of the verification that every linear
combination of these brackets is still a Poisson bracket on $e(3)$
and $so(4)$ respectively.

This Proposition allows us to say that the mapping $\widehat{f}$
(\ref{st-pM}) is a twisted Poisson map, which identifies two
bi-Hamiltonian manifolds $ e(3)$ and $ so(4)$ such that
\[\begin{array}{c}
\xymatrix{ {\mathscr P}_1  \ar[dr]  & \ar[dl]{\mathscr P}^*_1
\\
{\mathscr P}_2 \ar[ur] & \ar[ul]{\mathscr P}^*_2 }
\end{array}\qquad\mbox{\rm instead of}\qquad
\begin{array}{c} \xymatrix{ {\mathscr P}_1  \ar[r]  &
\ar[l]{\mathscr P}_1^*
\\
{\mathscr P}_2 \ar[r] & \ar[l]{\mathscr P}_2^*
}\end{array}\quad\mbox{\rm as for the  usual Poisson map.}
\]
The bi-Hamiltonian systems are defined on the bi-Hamiltonian
manifolds by the coefficients of the Casimir functions of the
Poisson pencil $\{.\,,.\}_\lambda=\{.\,,.\}_1-\lambda\{.\,,.\}_2$.
In our case on the bi-Hamiltonian manifold $e(3)$ cubic polynomial
$P_3(\lambda)=|\ell|^2$ (\ref{P3-K})
\bq\label{P3-e3}
P_3(\lambda)=\sum_{k=0}^3 \mathcal H_k\lambda^k=\mathcal
A\lambda^3+(\mathcal B-\mathcal A\mbox{\rm tr}\,\mathbf
A)\lambda^2+\dfrac{H_1}4\,\lambda+\dfrac{H_2}4\,.
\eq
is a Casimir of the Poisson pencil
\bq\label{poi-e3}
 \mathscr
P_\lambda=\mathscr P_2-\lambda \mathscr P_1,\qquad \mathscr
P_\lambda\,dP_3(\lambda,p,M)=0\,.
\eq
Here $\mathscr P_1$ and $\mathscr P_2$ are the Poisson tensors
associated with the brackets $\{.,.\}_1$ (\ref{new-e3}) and
$\{.,.\}_2$ (\ref{sB-e3}).

On the bi-Hamiltonian manifold $so(4)$ polynomial
\bq\label{P3-o4}
\widehat{P}_3(\lambda)=\widehat{f}\,\Bigl(P_3(\lambda)\Bigr)=\sum_{k=0}^3
\widehat{\mathcal H}_k\lambda^k=\mathcal
A^*\,\lambda^3+\widehat{H}_1\lambda^2+\widehat{H}_2\,\lambda-\det\mathbf
A\,\mathcal B^*\,
\eq
is a Casimir of the Poisson pencil
\bq\label{poi-o4}
 {\mathscr
P}^*_\lambda= {\mathscr P}_1^*-\lambda {\mathscr P}^*_2,\qquad
{\mathscr P}^*_\lambda\,d\widehat{P}_3(\lambda,s,t)=0\,.
\eq
Here ${\mathscr P}_1^*$ and ${\mathscr P}_2^*$ are the Poisson
tensors associated with the brackets $\{.,.\}^*_1$ (\ref{2o3}) and
$\{.,.\}^*_2$ (\ref{sB-o4}).

As usual, coefficients of the polynomial $P_3(\lambda)$
(\ref{P3-e3}) form a bi-Hamiltonian hierarchy on $e(3)$ starting
from a Casimir $\mathcal H_0$ of $\mathscr P_2$ and terminating
with a Casimir  of $\mathscr P_1$
\[
\mathscr P_2 d\mathcal H_0=0,\qquad \mathscr P_2d\mathcal
H_{i+1}=\mathscr P_1d\mathcal H_i,\qquad \mathscr P_1d\mathcal
H_3=0\,.
\]
Coefficients of the polynomial $\widehat{P}_3(\lambda)$ (\ref{P3-o4})
form the similar Lenard chain on $so(4)$. Therefore, the
Steklov-Lyapunov system on $e(3)$ and the Steklov system on $so(4)$
are the Gelfand-Zakharevich systems \cite{gz00}.

It is known that the Gelfand-Zakharevich systems admit the
St\"ackel separation of variables (see \cite{ped02} and references
within). However in our case tensors $\mathscr P_{1,2}$ and
${\mathscr P}^*_{1,2}$ are degenerate and, therefore, the
corresponding separated variables could be obtained after a
suitable reduction of the Poisson structures \cite{ped02,mt00}
only.

A natural way to do that is to fix the values of the Casimir
functions of $\mathscr P_{1}$ or ${\mathscr P}_{1}^*$. However, in
the our case each Poisson tensor $\mathscr P_1$ ($\mathscr P_1^*$) or
$\mathscr P_2$ ($\mathscr P_2^*$) can be properly restricted to a
corresponding symplectic leaf, but the other tensor does not restrict
to same leaf. So, a quite general reduction technique given by the
Marsden-Ratiu theorem \cite{ped02,mt00} have to be applied in this
case.

In fact we  make the necessary reduction  when we introduce the
chart $(x,J)$ related to $(p,M)$ by the map $\widetilde{f}_g$
(\ref{fg}). Namely, composition of the maps $\widehat{f}$
(\ref{st-pM}) and $\widetilde{f}_g$ (\ref{fg})  gives rise to the
second Poisson brackets on the second sample of $e(3)$ (\ref{e3})
\bq\label{22e3}
\{J_i,J_j\}'_2=\varepsilon_{ijk} a_kJ_k,\qquad
\{J_i,x_j\}'_2=\varepsilon_{ijk} a_jx_k-\mathbf
C_{ii}\dfrac{x_1x_2x_3}{|x|^2}\dfrac{x_j}{x_i},\qquad\{x_i,x_j\}'_2=0\,.
\eq
The Casimir of the  Poisson pencil ${\mathscr
P}'_{1}-\lambda{\mathscr P}'_{2}$  coincides with the cubic
polynomial $\widetilde{P}_3(\lambda)$ (\ref{3P3})
\[ \widetilde{P}_3(\lambda)=\delta\lambda^3+\gamma\lambda^2
+\widetilde{H}_1\lambda+\widetilde{H}_2\,,\qquad{\mathscr
P}'_\lambda\,d\widetilde{P}_3(\lambda,x,J)=0\,.
\]
This polynomial Casimir includes two numerical constants $\gamma$ and
$\delta$ instead of the Casimir functions and, therefore, the
corresponding symplectic foliation is trivial. According to
\cite{ts01}, in this case the  separated variables (\ref{ell-S}) are
the solutions of the following system of algebraic equations
\[
\dfrac{\partial }{\partial
\gamma}\,\widetilde{P}_3(\lambda,x,J)=0\,,\qquad\dfrac{\partial
}{\partial \delta}\,\widetilde{P}_3(\lambda,x,J)=0\,.
\]
These separated variables have to coincide with the eigenvalues of
the reduced Nijhenuis tensors $\mathscr N=\mathscr P_1\mathscr
P_2^{-1}$ or $\widehat{\mathscr N}=\widehat{\mathscr
P}_1\widehat{\mathscr P}_2^{-1}$. As sequence these variables have
to be in the involution with respect to both Poisson brackets
similar to the integrals of motion. We can check this property of
the K\"otter separated variables (\ref{sep-K}) without reduction
of the Poisson tensors.

\begin{prop}
The vectors
\[x=\dfrac{(M-\mathbf B p)\times  p}{|(M-\mathbf B p)\times
p\,|}\,\qquad\mbox{\it and}\qquad \widehat{x}=\dfrac{\sqrt{\mathbf
A^\vee}t\times s}{\sqrt{\mathbf A^\vee}t\times s},
\]
and, therefore,  zeroes of the standard equations
\[
\bigl\langle x,(\lambda-\mathbf
A)^{-1}x\,\bigr\rangle=0,\,\qquad\mbox{\it and}\qquad \bigl\langle
\widehat{x},(\lambda-\mathbf A)^{-1}\widehat{x}\,\bigr\rangle=0\]
are in the involution  with respect to both Poisson brackets on
$e(3)$ and $so(4)$ respectively.
\end{prop}
This Proposition has a straightforward consequence. The roots
$\widehat{u}_{1,2}=\widehat{f}(u_{1,2})$ of the non-diagonal entry
of the corresponding rational Lax matrix (\ref{lax-G})
\bq
\widehat{e}(\lambda)=\dfrac{(\lambda-\widehat{u}_1)(\lambda-\widehat{u}_2)}{\det(\lambda
-\mathbf A)}
=\sum_{i=1}^{n=3}\dfrac{\widehat{x}_i^2}{\lambda-a_i}=0
\eq
are in the involution
$\{\widehat{u}_1,\widehat{u}_2\}_1^*=\{\widehat{u}_1,\widehat{u}_2\}_2^*=0$
on $so(4)$ and, according to Proposition \ref{pr-bob}, they
satisfy to the following equations
\bq\label{evsep-St}
\dfrac{d}{dt}{\widehat{u}}_1=\dfrac{2\sqrt{\widehat{P}_3(\widehat{u}_1)
\det(\mathbf A-\widehat{u}_1)\,}}{\widehat{u}_1-\widehat{u}_2},\qquad
\dfrac{d}{dt}{\widehat{u}}_2=-\dfrac{2\sqrt{\widehat{P}_3(\widehat{u}_2)
\det(\mathbf A-\widehat{u}_2)\,}}{\widehat{u}_1-\widehat{u}_2}\,.
\eq
Hence $\widehat{u}_{1,2}$ are the separated variables for the
Steklov system on $so(4)$. These variables are poles of the
eigenvector $\widehat{\boldsymbol \Psi}(\lambda)$ of the Lax
matrix ${\widehat{\mathscr L}}_e(\lambda)$
\[
{\widehat{\mathscr
L}}_e(\lambda)_{ij}=\varepsilon_{ijk}\sqrt{\lambda-a_k}\left(s_k+\lambda^{-1}\sqrt{\mathbf
A^\vee_k}t_k\right)
\]
with the following dynamical normalization
\[
\widehat{\boldsymbol \alpha}=|\sqrt{\mathbf A^\vee\,}\, t\times
s\,|^{-1}\,\mathbf W\,s.
\]
The conjugated momenta $\widehat{v}_{1,2}$ are easy defined by the
diagonal entry of the corresponding rational Lax matrix
(\ref{lax-G})
\[
\widehat{v}_j=\left.
-\dfrac{1}{2\widehat{u}_j}\,\{\widehat{H}_1,\widehat{e}(\lambda)
\}\right|_{\lambda=\widehat{u}_j}\,,
\qquad\{\widehat{v}_j,\widehat{u}_j\}^\ast_1=\delta_{jk}
\]
after calculation of the  relation (\ref{rmat-rel}) with respect to
the second Poisson brackets (\ref{22e3}). The corresponding separated
equations are equal to
\[
\widehat{v}_j^2+\dfrac{\widehat{P}_3(\widehat{u}_j)}{\widehat{u}_j^2\,\det(\widehat{u}_j
-\mathbf A)}=0\,,\qquad j=1,2,
\]
such that the St\"ackel matrix
\bq\label{St2-mat}
\widehat{\mathcal S}=\left(%
\begin{array}{cc}
  u_1^2 & u_2^2 \\
  u_1 & u_2 \\
\end{array}%
\right)\,
\eq
is another block of the same Brill-Noether matrix $\mathcal
U_{\,\mathcal C}$ (\ref{BN-mat}). According to \cite{ts99b} the
St\"ackel systems associated with the different blocks of a common
Brill-Noether matrix are related by canonical transformation of the
time.

In contrast with the Steklov-Lyapunov system for the St\"ackel matrix
(\ref{St2-mat}) associated with the Steklov system we do not know how
to introduce coordinates in which equations of motion are the Newton
equations.

\section{Concluding remarks}
The first result in this paper is that, starting with the K\"otter
separated variables we construct the Poisson map which transforms
the Steklov-Lyapunov case of the Kirchhoff equations to the Newton
equations on the sphere. A natural way to construct such maps is
to identify the separated variables and the corresponding
separated equations.

The separation of variables for the Steklov-Lyapunov system on
$e(3)$ and for the Steklov system on $so(4)$ is discussed in
framework of the Sklyanin method and in the bi-Hamiltonian
approach. The main unsolved questions are how to construct a
suitable normalization of the Baker-Akhiezer function and a
suitable reduction of the degenerate Poisson tensors. To solve
these questions on this example we could to construct unknown
separated variables for the Clebsch system on $e(3)$ and the
Shottky-Manakov system on $so(4)$. As above, the bi-Hamiltonian
structure of these manifolds are defined by the similar linear
change of variables proposed in \cite{bob83}.

As a last remark we recall that in \cite{fed03} there are
separated variables, the elliptic and rational Lax matrices and
the compatible Poisson tensors for the multi-dimensional Steklov
systems. In our opinion these results deserve further
investigation in framework of the separation of variables theory
and of the general reduction theory for bi-Hamiltonian manifolds.

The research was partially supported by RFBR grant 02-01-00888.

\end{document}